\documentclass[twocolumn,resetfootnote]{aastex701}

\begin{document}

\title{New Online Database of Symbiotic Variables: Catalog and Statistical Overview of Symbiotic Binaries}

\author[orcid=0000-0001-6355-2468,sname='Jaroslav Merc']{Jaroslav Merc}
\affiliation{Astronomical Institute, Faculty of Mathematics and Physics, Charles University, 
V Hole\v{s}ovi\v{c}k{\'a}ch 2, 180 000, Prague, Czech Republic}
\affiliation{Instituto de Astrof\'isica de Canarias, Calle Vía Láctea, s/n, E-38205 La Laguna, Tenerife, Spain}
\email[show]{jaroslav.merc@mff.cuni.cz} 

\author[orcid=0000-0003-4299-6419,sname='Rudolf G{\'a}lis']{Rudolf G{\'a}lis}
\affiliation{Institute of Physics, Faculty of Science, P. J. \v{S}af{\'a}rik University, 
Park Angelinum 9, 040 01 Ko\v{s}ice, Slovak Republic}
\email[]{rudolf.galis@upjs.sk}

\author[orcid=0000-0002-4387-6358,sname='Marek Wolf']{Marek Wolf}
\affiliation{Astronomical Institute, Faculty of Mathematics and Physics, Charles University, 
V Hole\v{s}ovi\v{c}k{\'a}ch 2, 180 000, Prague, Czech Republic}
\email[]{marek.wolf@matfyz.cuni.cz} 

\begin{abstract}
We present the New Online Database of Symbiotic Variables (NODSV), a comprehensive and publicly accessible catalog of known and candidate symbiotic stars in the Milky Way and nearby galaxies. The database provides an up-to-date census of confirmed symbiotic binaries and systematically compiles information previously scattered across the literature, including photometric and spectroscopic properties, orbital parameters, and characteristics of both their cool and hot stellar components. It further records auxiliary diagnostics such as detected emission lines, flickering, X-ray emission, jets, or information about outburst activity. In its current release, NODSV contains nearly 1\,400 objects, classified into confirmed symbiotic stars, three categories of candidates, and misidentified sources. Based on the collected data, though originating from heterogeneous studies, we present a statistical overview of the confirmed symbiotic population, highlighting the distributions of orbital parameters and the properties of the cool giants and their hot companions. Designed as a dynamic and evolving resource, NODSV provides a foundation for future observational campaigns and theoretical investigations of symbiotic binaries.
\end{abstract}



\section{Introduction}
Symbiotic stars are long-period interacting binaries in which a cool evolved giant transfers matter to a hot compact companion, usually a white dwarf or, more rarely, a neutron star. The mass transfer, either via wind or Roche-lobe overflow, or a combination of those (wind Roche-lobe overflow), creates a complex environment of accretion structures, ionized gas, and/or dust. These systems exhibit a remarkable diversity of photometric and spectroscopic behaviours, and serve as valuable laboratories for studying mass transfer in late stellar evolution, with possible links to Type Ia supernova progenitors \citep[see reviews on symbiotic stars by][and references therein]{2012BaltA..21....5M,2019arXiv190901389M,2025Galax..13...49M}.

The peculiar characteristics of symbiotic stars were recognized in the early 20th century \citep[e.g.,][]{1912AnHar..56..165F,1920HarCi.221....1C}, when they were described as “stars with combination spectra" due to the simultaneous presence of low-temperature features (e.g. TiO bands) and high-temperature indicators (e.g. the He II emission line at 4\,686 \AA{}). The term “symbiotic stars" was introduced later by Merrill during a presentation at the American Astronomical Society meeting at Yerkes Observatory in 1941 \citep{1958LIACo...8..436M}.

Initially, research focused on individual systems. The first tabulated list was compiled by \citet{1954ApJS....1..175B}, who reported 23 “stars with combination spectra" in his catalog of late-type emission-line stars. A~few years later, \citet{1957gano.book.....G} expanded the group to 32 objects, while \citet{1969CoKon..65..395B} listed 21 confirmed and 16 suspected symbiotic stars. Classification at the time relied mainly on spectroscopic appearance, the cool continuum combined with strong emission lines, and/or distinctive photometric behavior such as nova-like outbursts.

The first comprehensive compilations of symbiotic stars appeared in the 1980s. The catalog of \citet{1984PASAu...5..369A} listed 129 confirmed and 15 suspected systems, including six in external galaxies. Shortly after, \citet{1986syst.book.....K} compiled 133 confirmed and 20 suspected symbiotics, while \citet{1988BAFOE..45...13V,1991BAFOE..55...21V} expanded the tally to 179 objects. These efforts culminated in the widely used catalog of \citet{2000A&AS..146..407B}, which contained 188 confirmed systems and 30 candidates. Since then, the number of confirmed symbiotics has more than doubled, driven by systematic searches both in the Milky Way \citep[see, e.g., ][]{2008A&A...480..409C,2010A&A...509A..41C,2013MNRAS.432.3186M,2014MNRAS.440.1410M,2014A&A...567A..49R,2021MNRAS.505.6121M,2021MNRAS.502.2513A,2023MNRAS.519.6044A,2024ApJ...962..126X,2025MNRAS.543.2292L,2025ApJ...987..147C,2025ApJ...995...14Z,2025OJAp....8E.122B,2026MNRAS.545f2146M,Merc+Gaia, 2026arXiv260424730M,2026MNRAS.546ag105A,2026A&A...708A..28C} and in nearby galaxies \citep[e.g.,][]{2014MNRAS.444..586M,2017MNRAS.465.1699M,2018arXiv181106696I}. Many additional candidates have been identified in large-scale photometric and spectroscopic surveys, but information on these systems remains scattered throughout the literature.

The only more recent census was presented by \citet{2019ApJS..240...21A}, who combined data from 2MASS, WISE, and \textit{Gaia} for 323 confirmed and 87 candidate systems drawn from the literature. Their study focused on infrared classification, \textit{Gaia} DR2 properties, the occurrence of Raman-scattered O VI lines, and X-ray detections. However, no up-to-date resource has existed that consolidates all available multi-wavelength, photometric, spectroscopic, and classification information for both confirmed and candidate symbiotic stars in a single, consistent, and accessible form.

To address this gap, contemporaneously with \citet{2019ApJS..240...21A}, we introduced the first version of the \textit{New Online Database of Symbiotic Variables} \citep[NODSV;][]{2019RNAAS...3...28M,2019AN....340..598M}. This resource compiles published information for confirmed and candidate systems in both the Milky Way and external galaxies, including multi-band photometry, spectroscopic information, orbital and stellar parameters, and classification notes. The database is regularly updated as new discoveries are reported, and the present work describes its advanced version in detail.

The paper is organized as follows. Section~\ref{sec:database} describes the NODSV, including its content and structure. In Section~\ref{sec:population}, we present the current Galactic and extragalactic symbiotic star populations included in the NODSV, with a discussion of orbital parameters and properties of their cool and hot components summarized in Sections~\ref{sec:orbital}, \ref{sec:cool}, and \ref{sec:hot}, respectively. Finally, based on the data compiled in the database, we summarize the typical parameters of symbiotic stars in Section~\ref{sec:conclusions}.

\section{New Online Database of Symbiotic Variables}\label{sec:database}
The steep increase in the number of known symbiotic stars in recent years has created a pressing need for a~modern, comprehensive resource. Such a database is essential not only for detailed population studies but also as a robust input for machine-learning algorithms and other automated classification tools.

When we released the first version of the NODSV in 2019, the most recent used catalog \citep{2000A&AS..146..407B} was already two decades old. We therefore set out to create an up-to-date, unified repository that compiles consistent, well-referenced information for all known symbiotic systems. Beyond a static list, our goal was to build an interactive web portal that allows researchers to access, filter, and download data with ease (see below).

Unlike traditional printed or static online catalogues, which are often outdated by the time they appear, the NODSV is continuously maintained. New symbiotic systems can be added immediately upon discovery, and existing entries can be updated as soon as improved measurements or classifications become available. This ensures that the astronomical community has continual access to up-to-date lists of symbiotic variables, along with detailed information on individual objects.

\subsection{Classification criteria}
The aim of the NODSV is to compile all objects that have been classified as symbiotic stars or candidates in the literature or recorded in astronomical databases. Because the exact definitions of symbiotic binaries differ across sources, and some objects have been labelled as symbiotic without satisfying any widely accepted criteria, we do not simply adopt the published status. Instead, we evaluate the available observational evidence for each source and assign a classification in a consistent manner. {For this work, we adopt the physical definition of \citet{2013A&A...559A...6L} and \citet{2016MNRAS.461L...1M}, considering a binary to be a symbiotic star if it hosts an evolved red giant donor transferring mass to a hot degenerate companion (white dwarf or neutron star) and if this interaction produces observable effects at some wavelength (not necessarily in the optical). Consequently,} the NODSV includes both shell-burning systems with luminous white dwarf accretors and accreting-only systems in which the compact companion is a white dwarf or a neutron star. Interacting binaries with main-sequence accretors, or with donors that are not cool evolved stars, are excluded \citep[see discussion in][]{2025Galax..13...49M}.

An object is classified as a “Confirmed" symbiotic star only when the observations clearly show the presence of a cool evolved star, for example, through late-type giant absorption features in the optical or infrared spectra or the detection of Mira-type pulsations, together with evidence for a hot degenerate companion that is accreting matter from the donor. The interaction between the components must be evident at some wavelength, which in shell-burning white dwarf systems is often indicated by strong ultraviolet–optical emission lines with an ionization potential of at least 35 eV \citep[see][]{2000A&AS..146..407B,2013MNRAS.432.3186M}, and in accreting-only systems by signatures such as a strong UV excess, rapid optical/UV variability (flickering), or hard X-ray emission. In the latter case, only systems in which the observations rule out a main-sequence accretor are considered confirmed. Objects showing Raman-scattered \mbox{O VI} emission lines {\citep[][]{1989A&A...211L..31S}} together with He II 4\,686 \AA{} are also treated as confirmed, even if the presence of the cool giant is not directly evident\footnote{\citet{2000A&AS..146..407B} suggested that the presence of Raman-scattered O VI lines alone is sufficient for a symbiotic classification. \citet{2019ApJS..240...21A} noted that in all known symbiotics with Raman-scattered O VI lines, the He II 4\,686 \AA{} line is also present, and that detections of Raman O VI without He II 4\,686 are dubious.}.

In the first version of the database \citep{2019RNAAS...3...28M,2019AN....340..598M}, the classification scheme distinguished only between confirmed symbiotic stars and candidates. With the growing number of entries in the NODSV, however, it became clear that a finer subdivision of candidates was needed to reflect the amount and quality of available information. In the current version, candidates are divided into three categories: “Likely", “Possible", and “Suspected". “Likely" objects are strong symbiotic candidates that do not meet all confirmation criteria but show multiple pieces of evidence pointing to a symbiotic nature, for instance, an optical spectrum revealing a cool evolved star with typical symbiotic photometric variability but only low-ionization emission lines, or an accreting-only candidate with UV excess and flickering where a main-sequence accretor cannot be excluded. “Possible" candidates are those for which some observational material exists and does not favor any alternative classification, but further data are required before they can be fully confirmed, for example, an optical spectrum showing some emission lines with the presence of a cool evolved star supported by infrared brightness. “Suspected" candidates comprise all remaining objects classified as symbiotic in the literature on the basis of indirect indicators such as symbiotic-like variability, characteristic colors, or a possible association of X-ray emission with a cool evolved star.

In addition to confirmed and candidate systems, the NODSV also maintains a list of “Misclassified" objects, meaning sources that were once considered symbiotic (either confirmed or candidates) but have subsequently been shown to be of a different nature. This is intended to prevent outdated classifications from re-entering the literature. Examples include DT Ser, {long considered a possible symbiotic star, later even} suggested to be a symbiotic star surrounded by a planetary nebula \citep{2013A&A...558A...2M}, but later demonstrated to be merely a superposition of unrelated objects\footnote{{Both the giant and the assumed ionizing central star of the PN are detected by \textit{Gaia} \citep[][]{2016A&A...595A...1G,2023A&A...674A...1G}, separated by $\sim$5\arcsec{}, with parallaxes of 1.97$\pm$0.24 mas and 0.12$\pm$0.08 mas, respectively, and inconsistent proper motions, confirming they are unrelated.}} \citep{2014MNRAS.445.1605F}. Despite this, it still appears as a possible symbiotic star in the catalog of \citet{2019ApJS..240...21A}. Another case is Hen 2-379, classified as a possible symbiotic by \citet{2000A&AS..146..407B} on the basis of a cool G–K giant spectrum and emission lines with ionization potentials up to 35.1 eV. The same study, however, questioned whether the giant and the nebula were physically associated. This superposition was later confirmed by \citet{2013MNRAS.432.3186M}, leading to the reclassification of the object as a planetary nebula. Yet this correction is not reflected in the catalog of \citet{2019ApJS..240...21A}. Similarly, IGR J16393-4643 was proposed as a symbiotic system by \citet{2010A&A...516A..94N}, but the association with a red giant donor was rejected by \citet{2012ApJ...751..113B}. The system has even been claimed to have an orbital period of $\sim$4.2 days \citep{2010ATel.2570....1C}, which is incompatible with a symbiotic classification. Other examples include WY Vel, recognized as a VV Cep star by \citet{1973ApJ...185..899S}, but nevertheless included as a confirmed symbiotic in earlier catalogs \citep{1957gano.book.....G,1988BAFOE..45...13V} and still flagged as such in SIMBAD; and Hen 3–814, which SIMBAD lists as a confirmed symbiotic star with a reference to \citet{1976ApJS...30..491H}, although that work in fact identified it as a VV Cep binary, a classification later confirmed by more detailed analyses \citep[e.g.,][]{2003A&A...397..927P}.

\subsection{Structure of the database}

The NODSV is organized into two main parts: Galactic symbiotic stars (confirmed and candidate systems) and their extragalactic counterparts. The Galactic part currently contains over a thousand entries (excluding those reclassified as misidentified), of which nearly 300 are confirmed symbiotics\footnote{Please refer to the online version for the most up-to-date numbers: \url{https://sirrah.troja.mff.cuni.cz/~merc/nodsv/}.}. The extragalactic section holds almost two hundred objects distributed across 16 galaxies, including the Magellanic Clouds, M31, M33, and several dwarf spheroidals. Among these, 71 are confirmed symbiotic binaries. A detailed breakdown of the current numbers in each galaxy is provided in Table~\ref{tab:numbers}. 

\begin{table}
\centering
\caption{Summary of objects in the NODSV by host galaxy and classification. Columns C, L, P, S, and M denote “Confirmed” symbiotic stars, “Likely,” “Possible,” and “Suspected” candidates, and “Misclassified” objects, respectively.}\label{tab:numbers}
\begin{tabular}{lcccccc}
\hline
Galaxy & C & L & P & S & M & Total \\
\hline
Milky Way       & 284 & 46 & 78 & 627 & 155 & 1190 \\
Draco Dwarf     & 1   & 0  & 3  & 0   & 0   & 4    \\
IC10            & 1   & 0  & 0  & 0   & 0   & 1    \\
LMC             & 10  & 0  & 5  & 24  & 3   & 42   \\
M31             & 32  & 5  & 3  & 8   & 1   & 49   \\
M33             & 12  & 0  & 0  & 0   & 1   & 13   \\
M81             & 0   & 0  & 1  & 0   & 0   & 1    \\
M87             & 0   & 0  & 0  & 9   & 0   & 9    \\
NGC 185         & 1   & 0  & 0  & 0   & 0   & 1    \\
NGC 205         & 1   & 0  & 2  & 0   & 0   & 3    \\
NGC 300         & 0   & 2  & 0  & 8   & 0   & 10   \\
NGC 55          & 0   & 0  & 0  & 3   & 0   & 3    \\
NGC 6822        & 1   & 0  & 11 & 0   & 0   & 12   \\
SMC             & 12  & 1  & 7  & 5   & 1   & 26   \\
Sculptor Dwarf  & 0   & 0  & 0  & 9   & 0   & 9    \\
Willman 1       & 0   & 0  & 0  & 1   & 0   & 1    \\
\hline
\textbf{Total}  & 355 & 54 & 110 & 694 & 161 & 1374 \\
\hline
\end{tabular}
\end{table}

Users can interact with the database in two ways: either by downloading the complete dataset (or selected tables) for offline analysis or by exploring its content directly through the dedicated web portal. The portal provides both tabular access to the data and individual object pages for each symbiotic system.  

\subsubsection{Tables with data}

The database compiles an extensive set of observational and physical parameters. For each object, we provide positional information, photometry across multiple wavelength ranges, and observational diagnostics such as the presence of outbursts, extended nebulae, jets, flickering, X-ray or radio detections, and infrared classification. Orbital parameters are also included where available (e.g., periods, ephemerides, eclipses), as well as fundamental properties of the binary components (spectral types, effective temperatures, masses, radii, luminosities, pulsational behavior, etc.).  

Each entry is cross-referenced with other major databases (e.g., SIMBAD, VSX, GCVS) and survey identifiers from missions such as \textit{Gaia} DR3, 2MASS, \textit{WISE}, and \textit{IRAS}. Each entry is also crossmatched against the principal historical and recent catalogs of symbiotic binaries \citep{1954ApJS....1..175B,1957gano.book.....G,1969CoKon..65..395B,1984PASAu...5..369A,1986syst.book.....K,1988BAFOE..45...13V,1991BAFOE..55...21V,2000A&AS..146..407B,2019ApJS..240...21A}, with the NODSV indicating whether an object is listed there and under which designation.  

The tabular data are organized into eight thematic sections, which can be sorted, filtered, and compared directly in the web interface. From each table, the object name links to a dedicated page that provides detailed information, references, notes, and additional resources. All tables are also available for download in multiple formats for offline use.  

\subsubsection{Object pages}

Each symbiotic binary in the NODSV has an individual object page that consolidates the information contained in the tables. Every numerical value or property is accompanied by references linked directly to the SAO/NASA Astrophysics Data System\footnote{\url{https://ui.adsabs.harvard.edu/}}, allowing immediate access to the original sources. Notes on distinctive features of individual systems are included where relevant.  

In addition, the object pages provide direct links to external resources, such as SIMBAD, CDS, VSX, and survey portals, enabling the user to quickly navigate between the NODSV and complementary datasets. Together, the tables and object pages form a flexible environment for both quick look-ups and in-depth research.  

\section{Symbiotic population in the NODSV}\label{sec:population}
The considerable increase in the number of confirmed symbiotic stars (now exceeding 350 in the current NODSV release; Table \ref{tab:catalog}), together with the availability of new parameter estimates (orbital periods, spectral types, stellar masses, etc.), enables for a more systematic statistical analysis of the known population.

\begin{table*}[]
\centering
\caption{List of symbiotic stars and candidates in the current version of the NODSV. The table lists the NODSV designation, host galaxy, classification status (C, L, P, S, and M denote “Confirmed,” “Likely,” “Possible,” “Suspected,” and “Misclassified,” respectively), \textit{Gaia} DR3 identifier \citep[][]{2016A&A...595A...1G,2023A&A...674A...1G}, 2MASS identifier \citep{2006AJ....131.1163S}, equatorial coordinates, and \textit{Gaia} $G$ magnitude. Full, machine-readable table is available online. Up-to-date version of the NODSV, containing additional parameters, is available at the web address \url{https://sirrah.troja.mff.cuni.cz/~merc/nodsv/}.  }\label{tab:catalog}
\scriptsize
\begin{tabular}{rllcccccc}
\hline
No.  & Name     & Galaxy    & Status    & Gaia DR3..            & 2MASS J..            & Right ascension & Declination & \textit{Gaia G} \\
  &      &     &     &             &             & [HH MM SS] & [DD MM SS] & [mag] \\
\hline
1    & NGC 55 SySt-3 & NGC 55    & S & -                   & -                & 00 14 58.61                   & $-$39 11 59.14              & -            \\
2    & NGC 55 SySt-2 & NGC 55    & S & -                   & -                & 00 15 39.02                   & $-$39 14 40.60              & -            \\
3    & NGC 55 SySt-1 & NGC 55    & S & -                   & -                & 00 16 07.25                   & $-$39 16 31.91              & -            \\
...    & ... & ...    & ... & ...                   & ...                & ...                   & ...              & ...            \\
18   & SMC1          & SMC       & C & 4685137110311670784 & 00291086$-$7457398 & 00 29 10.83                   & $-$74 57 39.89              & 15.84        \\
19   & LIN 9         & SMC       & C & 4688455814367629440 & 00300739$-$7337190 & 00 30 07.39                   & $-$73 37 19.10              & 15.48        \\
...    & ... & ...    & ... & ...                   & ...                & ...                   & ...              & ...            \\
1085 & BF Cyg        & MW & C & 2038568676176360192 & 19235351+2940292 & 19 23 53.50                   & +29 40 29.08              & 9.67         \\
1086 & CH Cyg        & MW & C & 2130088038314195200 & 19243305+5014289 & 19 24 33.06                   & +50 14 28.87              & 5.37         \\
...    & ... & ...    & ... & ...                   & ...                & ...                   & ...              & ...            \\
1211 & Z And         & MW & C & 1941894322438077312 & 23333994+4849059 & 23 33 39.95                   & +48 49 05.93              & 9.13         \\
1212 & R Aqr         & MW & C & 2419576358847950592 & 23434939$-$1517043 & 23 43 49.49                   & $-$15 17 04.66              & 6.71         \\
1213 & CGCS 5926     & MW & S & 2016034975622911360 & 23454464+6252511 & 23 45 44.65                   & +62 52 51.11              & 12.94       \\
\hline
\end{tabular}
\end{table*}

\begin{figure*}[t]
\plotone{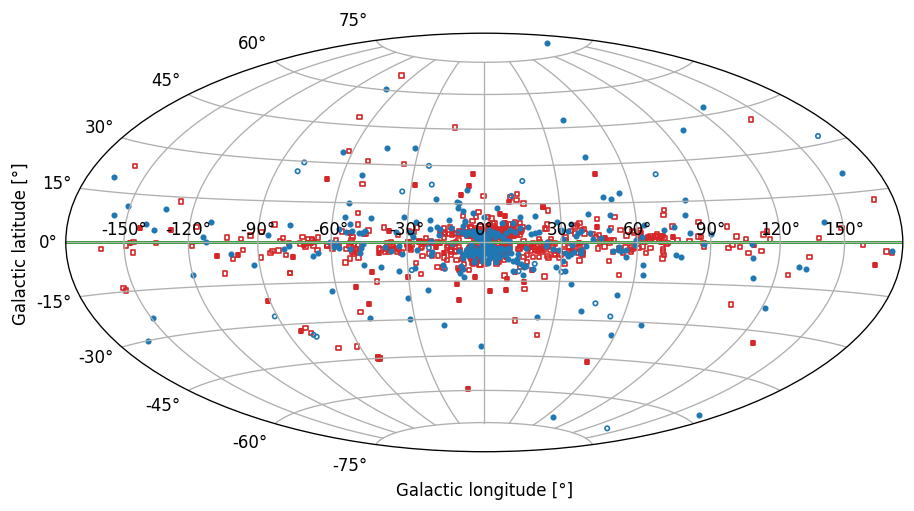}
\caption{Distribution of symbiotic stars in our Galaxy. Confirmed, likely, possible and suspected symbiotic stars are denoted by blue full dots, blue empty dots, red full squares, and red empty squares, respectively. Misclassified objects are not shown. The green line represents the equator of the Milky Way. The concentration towards the galactic plane/bulge is well visible.\vspace{8mm}
\label{fig:distribution_galactic}}
\end{figure*}

\subsection{Galactic symbiotic stars}

The present database contains almost 300 confirmed Galactic symbiotics, fewer than ten of which host neutron-star accretors. The remaining systems almost certainly contain white dwarfs as the hot components. Although the census of confirmed objects has grown substantially over the past two decades, it remains far below the expected Galactic population of symbiotic stars with white dwarf accretors, which has been estimated to lie between a few thousand and several hundred thousand \citep{1984PASAu...5..369A,1993ApJ...407L..81K,1992ApJ...397L..87M,2003ASPC..303..539M,2006MNRAS.372.1389L,2025A&A...698A.155L}. This discrepancy may partly reflect an overestimation of the fraction of red-giant–white-dwarf binaries that display symbiotic activity, but it is also likely that past surveys have systematically missed accreting-only symbiotics \citep[][see also the discussion on searches for new systems in \citeauthor{2025Galax..13...49M} \citeyear{2025Galax..13...49M}]{2016MNRAS.461L...1M}. Such systems, which lack strong emission lines but show signatures of accretion in the ultraviolet or X-rays, may constitute a large, yet largely unrecognized, fraction of the population. Caution is also required when comparing the observed sample with population synthesis predictions, since the NODSV includes weakly interacting systems (typically showing no or very weak emission lines), whose hot components have luminosities $\lesssim$10\,L$_\odot$ \citep[e.g., some objects from the sample of][]{2021MNRAS.505.6121M}. Such low-luminosity systems often fall below the selection thresholds adopted in synthetic population studies \citep[see, e.g.,][]{2022MNRAS.510.2707I}.

Including candidate systems, the current NODSV lists about 20 objects in which the hot component is probably a neutron star accretor, often referred to as symbiotic \mbox{X-ray} binaries. As with white-dwarf symbiotics, the number of known systems with neutron stars is much lower than predicted by population synthesis, which suggests between $\sim$50 and 1\,000 such binaries in the Milky Way \citep{2012MNRAS.424.2265L,2015AstL...41..114K,2019MNRAS.485..851Y}. This shortfall most likely reflects the intrinsic challenges of detecting them: they are typically inconspicuous at optical wavelengths, generally faint in the ultraviolet, and their \mbox{X-ray} luminosities often fall below the sensitivity limits of shallow X-ray surveys. Furthermore, they are expected to appear as X-ray sources only during a limited fraction of their lifetimes, when the accretion rate is high enough to produce detectable emission.

The distribution of Galactic symbiotic variables in Galactic coordinates is shown in Fig.~\ref{fig:distribution_galactic}. Almost all systems are concentrated near the Galactic plane, with $|b|<10^{\circ}$ containing about 80\% of the confirmed symbiotics and 75\% of the candidates in the current NODSV release. The fraction of confirmed systems increases to nearly 90\% when extending the latitude range to $|b|<18^{\circ}$. In Galactic longitude, the distribution reveals a strong concentration toward the bulge, with 59\% of confirmed systems and 57\% of candidates located within $330^{\circ}<l<30^{\circ}$. These fractions are affected by selection effects, since surveys have historically focused on regions close to the Galactic equator. Nevertheless, the concentration is also physical, reflecting both the higher stellar density in the Galactic disk and the fact that symbiotic stars are predominantly associated with the old bulge and thick-disk populations. 

\subsection{Extragalactic symbiotic stars}
The number of known extragalactic symbiotic stars and candidates has increased by an order of magnitude since the beginning of this century. A comprehensive analysis of the extragalactic symbiotic population, however, requires systematic searches, which have only recently begun and have so far concentrated mainly on M31 and M33 \citep{2014MNRAS.444..586M,2017MNRAS.465.1699M}. In other galaxies, symbiotics have typically been discovered serendipitously, most often as by-products of emission-line surveys \citep[e.g.,][]{2008MNRAS.391L..84G,2012MNRAS.419..854G,2015MNRAS.447..993G,2009MNRAS.395.1121K,2015A&A...574A.102S,2017MNRAS.464..739M,2018A&A...618A...3R}, through their X-ray emission \citep[e.g.,][]{2020MNRAS.499.3111S,2022MNRAS.512.5481S}, or via variability studies \citep[e.g.,][]{2016ApJS..227....1S}. A major advantage of studying extragalactic systems is that their distances are usually known with high accuracy, effectively corresponding to the well-established distances of their host galaxies. This allows reliable determination of intrinsic luminosities and other distance-dependent parameters, which in turn provides a solid basis for confronting observational data with theoretical models.

\section{Orbital parameters of known symbiotic stars}\label{sec:orbital}

Orbital periods of symbiotic stars have been determined either from photometric signatures (eclipses, reflection effects, ellipsoidal variability) or from spectroscopic monitoring. At the beginning of this century, periods were known for only about 20\% of the known systems: \citet{2000A&AS..146..407B} listed orbital periods for 34 confirmed symbiotics, including 20 with orbits derived from radial velocity curves, while \citet{2003ASPC..303....9M} compiled 27 spectroscopic solutions. Since then, further progress has come both from individual studies of particular objects and, more substantially, from systematic infrared spectroscopic monitoring, which provided many of the first spectroscopic orbits for red giant components \citep{2000AJ....120.3255F,2000AJ....119.1375F,2001AJ....121.2219F,2007AJ....133...17F,2008AJ....136..146F,2010AJ....139.1315F,2015AJ....150...48F,2017AJ....153...35F,2006ApJ...641..479H,2009ApJ...692.1360H,2019ApJ...872...43H}. Another major advance came from photometric surveys. The systematic analysis of long-term light curves by \citet{2013AcA....63..405G} yielded orbital periods for about 70 confirmed systems, including 34 first-time determinations.

\begin{figure}[t!]
\plotone{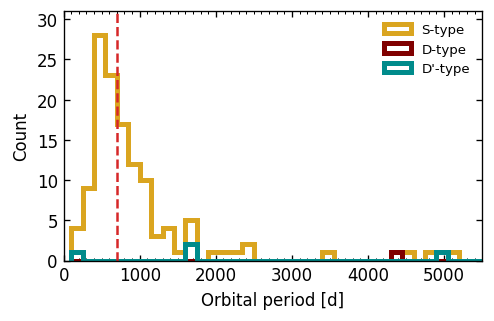}
\caption{Distribution of orbital periods for all confirmed symbiotic stars. S-, D-, and D'-type systems are shown in yellow, dark red, and cyan, respectively. A few objects with periods longer than 5\,500 days are not shown. The vertical red dashed line marks the median period of S-type systems (698 days).\vspace{8mm}
\label{fig:orbital_periods}}
\end{figure}

The NODSV now provides the most extensive compilation of orbital periods to date, with period estimates for 140 confirmed symbiotic stars ($\sim$39\% of the known Galactic and extragalactic population). Only 17 of these systems are extragalactic, and most of their periods are derived from photometric variability and remain uncertain. Draco C1 is the only extragalactic symbiotic with a~published spectroscopic orbit \citep{2020ApJ...900L..43L}, compared to nearly 50 Galactic systems. The majority ($\sim$ 90\%) of symbiotic stars whose orbital periods are below $\sim$900 days (S-type systems) have circular or nearly circular orbits (i.e., $e$ $<$ 0.15).

The reliability of period estimates depends strongly on the subtype. For S- and D'-type systems, orbital periods are typically determined from radial velocities or light curves. By contrast, most D-type systems are too wide for direct orbital solutions, and their periods are inferred indirectly from estimated separations or semi-major axes. The only exception so far is R~Aqr, for which radial velocity monitoring has provided a robust spectroscopic orbit with a period of 43.6 years \citep{2009A&A...495..931G}. All other D-type period estimates should therefore be regarded as highly uncertain. As an illustration, we consider V1016 Cyg. \citet{1988A&A...192L..10N} inferred an orbital period of approximately 9.5 years from variability in the O I and Mg II line fluxes observed in International Ultraviolet Explorer data. In contrast, \citet{1996A&A...310..211S} proposed a significantly longer period of about 80 years based on changes in the polarization angle. \citet{2000CoSka..30...99P} derived a period of roughly 15 years from long-term optical photometry, while \citet{2002ApJ...571..947B} suggested a timescale on the order of 544 years based on the angular separation of likely stellar components in Hubble Space Telescope imaging.

The distribution of orbital periods of confirmed symbiotic systems is shown in Fig. \ref{fig:orbital_periods}. Although this compilation represents the most extensive sample analyzed to date, it largely confirms the results obtained from earlier, smaller datasets \citep[e.g.,][]{1999A&AS..137..473M,2003ASPC..303....9M,2007BaltA..16....1M,2012BaltA..21....5M}. Most S-type symbiotics cluster in the range of 300--1\,100 days, with a maximum between 500 and 600 days. Approximately one-fifth of characterized S-type systems have periods longer than 1\,100 days, fewer than 5\% fall below 300 days, and only two have periods shorter than 200 days: V483 Sct \citep[197.6 days, albeit uncertain;][]{2001A&A...370..503M} and TX~CVn \citep[199 $\pm$ 3 days;][]{1989AJ.....97..194K}. All known D'-type systems have periods longer than 1\,500 days, with the exception of StH$\alpha$~190, for which \citet{2001A&A...369L...1M} reported a period of 171 $\pm$ 5 days based on photometry and preliminary radial velocities, while \citet{2001ApJ...556L..55S} suggested an even shorter value of 37--39 days from subsequent data. D-type systems hosting Mira variables are expected to have orbital periods of several decades to centuries ($\sim 10^{4}$--$10^{5}$ days), though reliable estimates exist for only a handful of cases.

The observed distribution contrasts sharply with predictions from population synthesis models. For instance, \citet{2006MNRAS.372.1389L} predicted a peak near $\sim$1\,500 days, with only about 20\% of symbiotics below 1\,000 days, essentially the opposite of what is observed. 

In about 7--10\% of confirmed symbiotic binaries in the NODSV, evidence for eclipses has been reported in the literature. This fraction is likely a lower limit, since long-term photometric monitoring is unavailable or has not yet been analyzed for several confirmed systems. In addition, prominent eclipses of the hot component and its surrounding ionized nebula by the cool giant are often visible only during active stages \citep[e.g.,][]{2011A&A...536A..27S,2012ApJ...750....5K,2022MNRAS.510.1404M}.

The incidence of eclipsing systems among symbiotics appears to be considerably higher than in binaries with comparable orbital periods in general \citep[cf. the Kepler samples in][]{2011AJ....141...83P,2016AJ....151...68K}. This difference can be understood in terms of geometry: in typical detached binaries with two stellar components, eclipses occur only at very high inclinations (near 90$^{\circ}$), whereas in symbiotics, particularly during outbursts, the eclipsed region is spatially extended. As a result, eclipses can be detected over a broader range of orbital inclinations.

\section{Cool components of symbiotic stars}\label{sec:cool}
By definition, the cool components of the symbiotic binaries are evolved giant stars. They dominate the optical and infrared spectra of the systems, and are therefore typically characterized using these wavelength regions. 

The largest uniform sample of the spectral types of the cool components in symbiotic binaries was published by \citet{1999A&AS..137..473M}, who classified the cool components in about 100 systems based on NIR spectra. \citet{2000A&AS..146..407B} collected the spectral types for almost all of the 188 confirmed symbiotic binaries in their catalog from the literature, though some of the classifications were limited only to distinguishing between spectral types G, K, or M. 

\begin{figure}[t]
\plotone{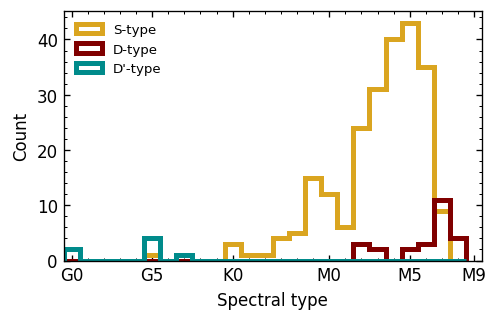}
\plotone{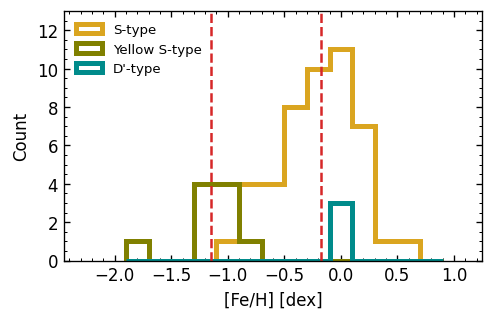}
\plotone{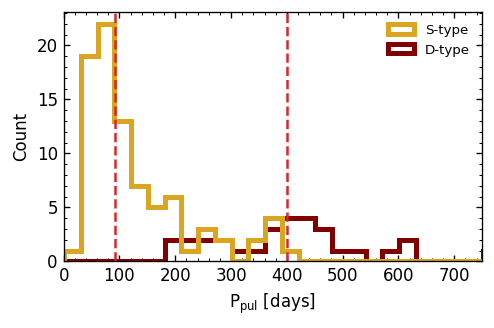}
\caption{Cool components of confirmed symbiotic stars.
\textbf{Upper panel:} Distribution of spectral types. S-, D-, and D'-type systems are shown in yellow, dark red, and cyan, respectively.
\textbf{Middle panel:} Distribution of metallicities of symbiotic giants. S-type systems with M-type giants are shown in yellow, K-type giants in olive, and D'-type systems (G giants) in cyan. Vertical red dashed lines indicate the median values for M-type giants ([Fe/H] = $-$0.18) and K-type giants ([Fe/H] = $-$1.15).
\textbf{Bottom panel:} Distribution of pulsation periods. S-type systems (typically SR or \mbox{OSARG} pulsations) are shown in yellow, and D-type systems (Mira pulsators) in dark red. Vertical red dashed lines denote median periods of 92.5 days for S-types and 400 days for D-types. \vspace{8mm}
\label{fig:spectral_types}}
\end{figure}

For the NODSV, we have compiled spectral types for 325 confirmed systems\footnote{Note that, as with other parameters in the NODSV, these values may have been derived using heterogeneous methods, since the database compiles results from the published literature.} (about 90\% of the whole sample) and for 178 candidate symbiotics (about 36\% of all candidates). This represents the largest collection of spectral types of symbiotic giants available to date. The distribution of spectral types for the confirmed systems is shown in the upper panel of Fig.~\ref{fig:spectral_types}. Our enlarged sample confirms earlier results \citep[e.g.,][]{1995AJ....109.1770M,1999A&AS..137..473M,2003ASPC..303....9M}: S-type symbiotics with M-type giants cluster strongly between M3 and M6, with a peak at M5. Interestingly, a secondary concentration is visible around K5--M0. This partly reflects the recent discoveries of new yellow symbiotics, but may also be influenced by classification challenges in this range, as objects with weak or absent TiO bands can be assigned later types than they truly are, potentially producing an artificial excess. Nevertheless, a mild overabundance of systems in this range was already noted by \citet{1995AJ....109.1770M}. In contrast, D-type symbiotics (hosting Mira variables) peak clearly at M7--M8, while the rare D'-type systems contain F/G-type giants, though their small number prevents meaningful statistical analysis.

Previous studies \citep{1980MNRAS.192..521A,1999A&AS..137..473M} noted that symbiotic giants tend to be of later spectral type than field giants in the solar neighborhood. Our larger sample, which now includes more yellow symbiotics and earlier M-type systems, does not change this picture: the later spectral types remain dominant, consistent with the idea that more evolved giants with larger radii and higher mass-loss rates are key ingredients for the symbiotic phenomenon \citep[][see also discussion in \citealt{2025A&A...695A..61M} and \citealt{2025arXiv250801304B}]{2003ASPC..303....9M}. \citet{1992A&A...255..171W} proposed that symbiotic giants are similar to those in the Galactic bulge rather than nearby bright giants, but this interpretation is challenged by more recent abundance studies.

While the metallicities of yellow symbiotic stars (with K giants) have been known since the 1990s \citep[see, e.g., ][and references therein]{2017ApJ...841...50P}, detailed chemical abundances of a large sample of M-type symbiotic giants became available only recently \citep[see, e.g., ][for earlier results]{1988A&A...198..179N}. Using high-resolution near-IR spectra and spectrum synthesis, \citet{2014MNRAS.440.3016M} and \citet{2015MNRAS.447..492G,2016MNRAS.455.1282G,2017MNRAS.466.2194G,2023MNRAS.526..918G} derived abundances for more than 50 symbiotic giants. The distribution of metallicities from the literature, compiled in the NODSV, confirms earlier indications \citep[e.g.,][]{2007BaltA..16....1M,2016MNRAS.455.1282G} that M giants in symbiotics are typically slightly sub-solar, with a median [Fe/H] = -0.18 (the middle panel of Fig.~\ref{fig:spectral_types}). This contrasts with the metal-rich, low-mass bulge M giants suggested by \citet{1992A&A...255..171W}. {It should be noted, however, that these abundance analyses rely on several simplifications (e.g., thin, plane-parallel, static, LTE atmospheres), which differ from the complex conditions in real red giant atmospheres; consequently, the derived abundances should be treated with caution \citep[see, e.g.,][]{2019arXiv190901389M}.} For completeness, yellow symbiotics studied so far host metal-poor ([Fe/H] $\approx$ -1.15), s-process enhanced K giants, whereas D'-type systems analyzed to date contain G giants with approximately solar metallicities.

\begin{figure}[t]
\plotone{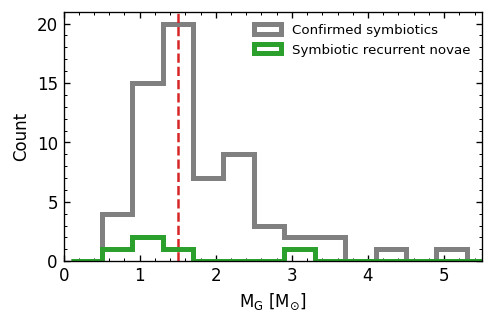}
\plotone{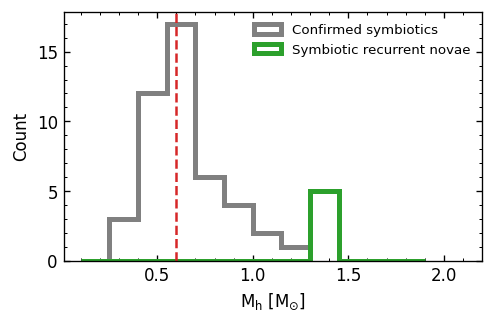}
\caption{Masses of binary components in confirmed symbiotic stars.
\textbf{Upper panel:} Masses of symbiotic giants (gray). The vertical red dashed line indicates the median value of 1.5~M${_\odot}$. Systems showing recurrent nova outbursts, including V407 Cyg, are shown in dark green. 
\textbf{Lower panel:} Distribution of symbiotic white dwarf masses (gray). The vertical red dashed line marks the median value of 0.6 M${\odot}$. Systems with recurrent nova outbursts are shown in dark green.\vspace{8mm}
\label{fig:masses}}
\end{figure}

Typical cool components of symbiotic stars occupy the region of the HR diagram corresponding to luminosity classes III--II, consistent with their identification as giants at the tip of the red giant branch (RGB) or early asymptotic giant branch (AGB), with a few systems among more evolved AGB stars. Mass estimates collected for symbiotic giants \citep{2007BaltA..16....1M,2010arXiv1011.5657M} indicate that the majority belong to the low-mass population, with $M_{\rm G} \lesssim 2$–3~M$_{\odot}$. This conclusion is reinforced by the larger sample of 66 masses of the cool components in the NODSV, where most giants have masses in the range 1–3~M$_{\odot}$\footnote{It should be emphasized that, for this parameter more than others, the input sources are highly heterogeneous and the methods used to estimate the masses vary between sources. Only in a very few cases are direct estimates available; more often, this is based on spectral types or isochrone fitting. Consequently, some values are significantly more reliable than others.} (upper panel of Fig. \ref{fig:masses}) and only a small fraction exceed 3~M$_{\odot}$, typically in D'-type systems. The median mass of the sample is 1.5~M$_{\odot}$.

Long-term photometric monitoring, particularly from large surveys, has revealed that a substantial fraction of symbiotic giants exhibit pulsations. Variability analyses of large samples \citep{2009AcA....59..169G,2013AcA....63..405G}, as well as individual studies, show that S-type systems usually display semi-regular pulsations or oscillate as OGLE small-amplitude red giants (OSARGs; \citealt{2004MNRAS.349.1059W}), while D-type systems host radial pulsators of Mira type.

In the NODSV, pulsation properties were compiled for 86 confirmed S-type symbiotic stars and 31 confirmed symbiotic Miras. The pulsation periods of S-type systems cluster strongly in the range 40–-200 days, with a median of 92.5 days, and only $\sim$15\% of S-type systems have periods exceeding 200 days (see lower panel of Fig. \ref{fig:spectral_types}). All D-type systems with Mira pulsators exhibit periods longer than 200 days, with a median of 400 days. Notably, the longest-period D-type pulsators are V407 Cyg ($\sim$745 days; \citealt{1990MNRAS.242..653M}), which underwent an outburst resembling a symbiotic recurrent nova, and the recently discovered symbiotic X-ray binary SRGA J181414.6$-$225604 ($\sim$1\,502 days; \citealt{2022ApJ...935...36D}).

\section{Hot components of known symbiotics}\label{sec:hot}
The hot components of symbiotic binaries are typically either luminous, hot white dwarfs in shell-burning systems, or faint accreting white dwarfs and, rarely, neutron stars in accreting-only systems. The number of symbiotic systems hosting neutron star accretors is currently too small for meaningful statistical analysis. Therefore, in this section, we focus on symbiotic stars containing white dwarfs. We have compiled several properties of these hot components from the literature for both confirmed symbiotic stars and candidates included in the NODSV.

In quiescence, the hot components generally contribute negligibly to the optical flux, and UV observations are not always available. Consequently, key parameters such as temperature and luminosity are often inferred indirectly, for example, from their effects on the surrounding nebula and from prominent emission lines {\citep[e.g.,][]{1981ASIC...69..517I,1988ASSL..145..107N,1991AJ....101..637K,1997A&A...327..191M,2016MNRAS.456.2558L,2020A&A...644A..49M}.}

\begin{figure}[t!]
\plotone{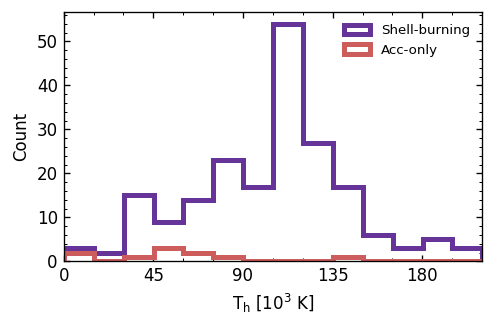}
\plotone{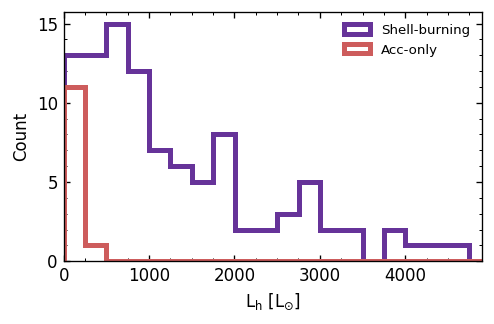}
\plotone{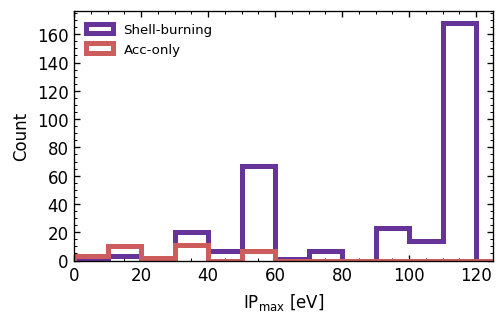}
\caption{Hot components of confirmed symbiotic stars. Shell-burning systems and accreting-only symbiotic stars are shown in purple and red, respectively.
\textbf{Upper panel:} Distribution of temperatures of symbiotic white dwarfs. \textbf{Middle panel:} Distribution of luminosities of white dwarfs in symbiotic systems. \textbf{Bottom panel:} Distribution of maximal ionization potential of emission lines in the quiescent spectra of symbiotic systems.\vspace{8mm}
\label{fig:hot}}
\end{figure}

White dwarfs in shell-burning symbiotic systems typically have temperatures of a few $\times 10^4$ to $10^5$ K and luminosities of $10^2$--$10^4$ L${_\odot}$ (upper and middle panels of Fig. \ref{fig:hot}). In accreting-only systems, the white dwarfs are cooler ($< 10^5$ K) and less luminous ($\sim 1$--$10^2$ L$_{\odot}$). These estimates should be treated with caution, as the temperatures are sometimes only lower limits derived from the maximum observed ionization potential \citep{1994A&A...282..586M}, very often from the presence of Raman-scattered O VI lines (with ionization potential of 114 eV; see lower panel of Fig. \ref{fig:hot}), while luminosities depend on distances, which are poorly constrained for many systems. Both temperature and luminosity can also vary with the activity state of the binary {\citep[see, e.g., ][]{1999A&A...347..478G,1994A&A...282..586M,2005A&A...440..995S}}. {The distributions are consistent with the results obtained earlier by, e.g., \citet{1991A&A...248..458M} or compiled by \citet{2003ASPC..303....9M}.}

Mass estimates for symbiotic white dwarfs were compiled for 17 and 22 systems by \citet{2007BaltA..16....1M} and \citet{2010arXiv1011.5657M}, respectively. In the NODSV, we now have masses for 50 confirmed systems. Most symbiotic white dwarfs have masses between 0.4 and 0.8~M$_{\odot}$, with a median of 0.6~M$_{\odot}$ (lower panel of Fig. \ref{fig:masses}), consistent with previous studies. Interestingly, the mass distribution closely resembles that of Galactic disk white dwarfs \citep[e.g.,][]{2008MNRAS.387.1693C,2013ApJS..204....5K,2015MNRAS.446.4078K,2016MNRAS.455.3413K}.

Neglecting any mass growth due to accretion from the cool companion, progenitor masses can be estimated from initial-final mass relations. Theoretical models \citep{2016ApJ...823..102C} suggest initial masses of $\sim$2--3~M$_{\odot}$ for a final mass of 0.6~M$_{\odot}$. Semi-empirical relations give slightly lower values: $\sim$2~M$_{\odot}$ \citep{2008MNRAS.387.1693C} and $\sim$1.4~M$_{\odot}$ \citep{2018ApJ...866...21C}. The latter may be underestimated, as the current symbiotic white dwarf must have been initially more massive than its companion, whose mass distribution peaks around 1.5~M$_{\odot}$.

A notable exception is a small group of symbiotic binaries exhibiting recurrent nova outbursts, whose white dwarfs have higher masses ($\sim$1.2--1.4~M$_{\odot}$). At least for RS Oph, there is evidence of the CO nature of the white dwarf \citep{2017ApJ...847...99M}. Since the theoretical upper mass limit for CO white dwarfs is \mbox{$\sim$1--1.1 M${_\odot}$} \citep[e.g.,][]{2013IAUS..281...36M}, these systems provide clear evidence that symbiotic white dwarfs can grow in mass, making them promising progenitors of Type Ia supernovae.

Finally, we also note that the NODSV includes information on emission lines in the optical and UV spectra of symbiotic stars, directly related to their hot companions, along with records of outburst activity, flickering \citep[see also][]{2024A&A...683A..84M}, jets, resolved nebulae, and detectable X-ray emission \citep[see also][]{2013A&A...559A...6L}.

\section{Conclusions}\label{sec:conclusions}

Symbiotic stars, with their rich variety of phenomena, serve as unique astrophysical laboratories for studying stellar evolution, mass transfer and accretion processes, stellar winds and their interactions, thermonuclear outbursts, jet formation and collimation, dust formation and destruction, and variable X-ray emission.

In this article, we presented the New Online Database of Symbiotic Variables, the most comprehensive and up-to-date catalog of symbiotic binaries. It compiles orbital, stellar, and observational parameters for all known symbiotic systems. Using these data, we can summarize the typical properties of symbiotic stars, while keeping in mind that the group is heterogeneous and some systems deviate significantly from any “prototype".

A typical S-type symbiotic binary ($\sim$77\% of all known symbiotics) has an orbital period of 300--800 days, a~nearly circular orbit, and consists of:

\begin{itemize}
\item a normal or bright M giant (spectral type M3--M6) semi-regularly pulsating with a period of 50--200 days, mass 1--2.5~M$_{\odot}$, slightly sub-solar metallicity, and mass-loss rate $\sim 10^{-7}$\,M$_{\odot}$\,yr$^{-1}$,
\item a hot white dwarf ($T_{\rm eff} > 10^5$ K) with mass \mbox{0.4--0.8~M$_{\odot}$}, either sustaining shell hydrogen burning on its surface ($L \approx 10^2$–$10^4$ L$_{\odot}$) or accreting-only with a lower luminosity ($L \approx 10^1$–$10^2$ L$_{\odot}$),
\item a compact circumbinary nebula formed from wind matter, particularly from the giant, with size of a~few au, temperature $\sim 10^4$ K, and electron density $n_e \sim 10^8$–$10^{12}$ cm$^{-3}$.
\end{itemize}

Some S-type systems deviate from this standard picture. Yellow symbiotic stars host metal-poor K giants instead of M giants, symbiotic recurrent novae contain more massive white dwarfs, and a few systems feature accreting-only neutron stars, which do not exhibit the typical symbiotic optical emission lines.

D-type symbiotic binaries have much longer orbital periods (tens of years) and consist of:
\begin{itemize}
\item a very evolved Mira (spectral type M7--M8) pulsating with a period of 250--600 days, surrounded by an optically thick dust envelope and losing mass at $\sim 10^{-5}$~M$_{\odot}$ yr$^{-1}$,
\item a low-mass white dwarf accreting from the giant’s wind,
\item a more extended (10--100 au), lower-density nebula with $n_e \sim 10^6$–$10^7$ cm$^{-3}$.
\end{itemize}

A small fraction of dusty systems host warmer G-type giants and are classified as D'-type symbiotics, comprising roughly 3\% of all known systems.

The NODSV provides a unified framework to study these diverse systems systematically. By making detailed stellar, orbital, and observational parameters readily accessible, the database facilitates statistical studies, comparisons across subtypes, and identification of rare or unusual systems. It is a powerful resource for the community that will support future observational campaigns, theoretical modeling, and machine-learning applications, ultimately advancing our understanding of binary evolution and the complex interactions that drive symbiotic activity.


\begin{acknowledgments}
We are grateful to the anonymous referee for the constructive and insightful comments that significantly improved the clarity and quality of this work. We thank all colleagues who have used the NODSV since its first release and provided valuable feedback, which has been essential for continuously improving this resource for the community. This research has made use of the SIMBAD database \citep[][]{2000A&AS..143....9W} and the VizieR catalogue access tool \citep{2000A&AS..143...23O}, CDS, Strasbourg Astronomical Observatory, France. 

JM was supported by the Czech Science Foundation (GACR) project no. 24-10608O. MW and JM were supported by the project COOPERATIO-PHYSICS of Charles University in Prague. RG was supported by the Slovak Research and Development Agency under contract No. APVV-24-0160.
\end{acknowledgments}

\software{astropy \citep{2013A&A...558A..33A,2018AJ....156..123A,2022ApJ...935..167A}, matplotlib \citep{Hunter:2007},
NumPy \citep{harris2020array}}





\bibliography{sample701}{}
\bibliographystyle{aasjournalv7}



\end{document}